\documentclass{moriond}
\usepackage{units}

\def\Journal#1#2#3#4{{#1} {\bf #2}, #3 (#4)}

\def\PRD{{\em Phys. Rev.} D}

\def\be{\begin{equation}}
\def\ee{\end{equation}}
\def\bea{\begin{eqnarray}}
\def\eea{\end{eqnarray}}

\begin{document}
\vspace*{4cm}
\title{Weakly modeled search for compact binary coalescences in the Einstein Telescope}

\author{A. Macquet$^1$, T. Dal Canton$^1$, T. Regimbau$^2$}

\address{$^1$Université Paris-Saclay, CNRS/IN2P3, IJCLab, 91405 Orsay, France\\
$^2$Univ. Savoie Mont Blanc, CNRS, Laboratoire d’Annecy de Physique des Particules - IN2P3, F-74000 Annecy, France}

\maketitle\abstracts{
We search for gravitational-wave (GW) signals from compact binary coalescences (CBC) in the $2024$ mock data challenge of the Einstein Telescope (ET) with a detection algorithm that does not rely on the waveform of the signal searched.
With the increased sensitivity of ET compared to current GW detectors, a very high rate of detectable sources is expected in the data, and the computational cost of the searches may become a limiting factor. This is why we explore the behavior of a weakly modeled search algorithm, which is intrinsically less sensitive than optimal search methods based on matched filtering techniques, but computationally much cheaper.
This search recovers a significant fraction of CBC signals present in the data: $38\%$ of the total number of binary black hole mergers, including $89\%$ of the systems with a total mass above \unit[$100$]{M$_\odot$}, as well as the majority of binary neutron star (BNS) mergers closer than \unit[$850$]{Mpc} ($z=0.17$). It is also able to estimate the chirp mass of the recovered BNS with an average precision of $1.3 \%$. 
We also find that the usual method for estimating the background in transient GW searches, that consists in time-shifting the data from one detector with respect to the others, is impacted by the presence of loud CBC signals in the data, so we use the null stream instead as a signal-free channel to estimate the background of this search.
}

\section{Introduction and motivation}

Expected in the mid-$2030$s, the Einstein Telescope (ET) \cite{ET} will be a ground-based gravitational wave (GW) detector which aims to increase the sensitivity by an order of magnitude compared to Advanced LIGO and Advanced Virgo in a frequency range between a few Hz to few kHz.
Its current planed design consists of three nested interferometers forming an equilateral triangle with \unit[$10$]{km} sides. Each detector will consist of two dual-recycled Fabry-Perot-Michelson interferometers tuned to be sensitive to high and low frequencies respectively, and the whole infrastructure will be built underground to limit the impact of seismic noise.

At its design sensitivity, ET should be able to detect most of the binary black holes mergers (BBH) with total mass in the range \unit[$10^1-10^3$]{M$_\odot$} up to a redshift $z\sim20$, and binary neutron stars mergers (BNS) up to $z\sim2$, allowing to probe the distribution of compact binary coalescences (CBC) at large redshift, among many other scientific objectives (a comprehensive description of the scientific targets of ET can be found in \cite{BB}).
In order to reach these objectives, current data analysis techniques used for LIGO/Virgo data must be adapted to deal with a significantly increased signal regime: the expected rate of detectable CBC in ET lies in the hundreds to thousands per day, compared to $\sim 1$ per day in Advanced LIGO at its design sensitivity.
In addition, the extended sensitivity at low frequency means that CBC signals will be present for a longer time in the frequency band of the detectors, up to several hours or days in the case of BNS.

Because CBC signals are well modeled and depend on a relatively small number of parameters ($16$ in the most general case), they can be searched nearly optimally in the data using matched filtering techniques, that consist in cross-correlating templates of the signals searched with the data.
Conversely, when a signal is not well enough modeled to be searched with matched filtering techniques, one must rely on weakly modeled search algorithms that make no or few assumptions on the exact waveform of the signal searched. Such methods typically rely on looking for patterns of excess power in some time-frequency representation of the data.
By construction, modeled searches are more sensitive to CBC signals than weakly modeled ones, but they are also much more computationally expensive because a large number of templates must be generated and cross-correlated with the data to cover the parameter space of CBC. In ET, that cost will further increase because of the large number of signals in the data, but also because of their extended duration, which will require much longer templates that are expensive to handle, up to a point that may become prohibitive. On the other hand, the improved sensitivity means that a large number of signals will have a high signal to noise ratio (SNR) and may therefore be detectable by weakly modeled searches. 

Here we study the ability of a weakly modeled search algorithm to recover CBC signals in ET. We run \texttt{PySTAMPAS}, a detection algorithm designed to search for long-duration ($\sim 1-1000$ s) transient GW signals \cite{pystampas}, on the 2024 ET mock data challenge (MDC). This work is described in more details in \cite{paper}.

\section{Dataset and methodology}

\subsection{ET mock data challenge}
The ET collaboration regularly issues MDCs in order to provide common datasets to test and demonstrate data analysis methods, and to precise the scientific objectives of the detector.
The MDC analyzed in this work simulates $31$ days of ET data for the $3$ nested detectors forming the triangle.
The data consist of Gaussian noise following the design sensitivity of ET, on top of which are injected $\sim 70000$ CBC signals following realistic astrophysical distributions compatible with the latest LVK observations \cite{GWTC-3}. Their distribution in total mass and redshift is represented in Fig. \ref{fig:M_vs_z}.

\subsection{Data analysis method}

\texttt{PySTAMPAS} is a data analysis pipeline originally designed to search for long-duration ($\sim 1-1000$ s) transient GW signals in LIGO/Virgo data with minimal assumptions on the morphology of the signal.
First, time-frequency maps (TF-maps) of the data are built by computing the short-time Fourier Transform of short segments of duration $0.5$, $1$, $2$, and $4$ s, which are then combined into a multi-resolution TF-map. Each map spans a duration of $512$ s and a frequency range $4-2000$ Hz.
Candidate GW signals (\textit{triggers}) are then identified by running a pattern recognition algorithm on the individual TF-maps. At this step, loose constraints may be imposed on the time-frequency morphology of the signal searched (hence the search being described as \textit{weakly} modeled).
Here we explore two complementary strategies. First, we run a \textit{seed-based} clustering algorithm that identifies bright pixels in the TF-map and groups them by proximity. This algorithms allows to reconstruct any signal morphology (provided it is contiguous in the time-frequency plane), but requires high SNR pixels. 
For long-lived BNS signals that may last up to several hours, the energy is spread over a large number of pixels in the TF-map, so individual pixels may not have a sufficiently high SNR to be picked up by the seed-based algorithm. Therefore we also use a \textit{seedless} clustering algorithm, whose principle is to fit pre-determined time-frequency templates on each TF-map. Here we fit Newtonian chirp templates, characterized by their chirp mass ($100$ values between $1$ and $5$ M $_\odot$) and coalescence time (every $0.5$ s).
%We do not take post-Newtonian terms and spin into account because we are working on the time-frequency space, and therefore the signal is binned in pixels of duration $0.25$ s $\times$ $0.5$ Hz. At this precision, the time-frequency morphology of the waveform is driven by the chirp mass 
As a by-product, the chirp mass of the template that maximizes the SNR can be used as a rough estimation of the actual chirp mass of the signal.

Finally, in order to discriminate true GW signals, that are expected to appear coherently in the data from the $3$ detectors, from noise-induced candidates, triggers extracted in single-detector TF-maps are then cross-correlated with coincident data from the $2$ other detectors, and a detection statistic $p_\Lambda$ is computed that reflects the overall coherence of each trigger across the $3$ detectors.

\section{Results}

\subsection{Background estimation}
To assess the significance of triggers found when analyzing the data, any given search must estimate the background distribution, \textit{i.e} the rate of triggers due to noise in the detector. Searches for transient GW signals usually estimate the background by shifting the data from one detector with respect to the others by an amount of time greater than the expected coherence time of the signal searched: coherent GW signals are therefore removed from the data while the statistical properties of the noise (which in general is neither Gaussian nor stationary) are preserved. Besides, this step can be repeated a large number of times with different values of the time shift to simulate an amount of background noise much larger than the actual duration of the dataset observed (up to hundreds or thousands of years).

We apply this method in this analysis at the cross-correlation stage: triggers found in each of the three detectors' data are time-shifted with respect to the data from the other two detectors before the computation of the coherent statistic is made. This process is repeated $1216$ times to generate $\sim$ 100 years of background, and the cumulative distribution of the ranking statistic $p_\Lambda$ can be used to set a threshold on the false-alarm rate (FAR) of the search.
However, we find that the background distribution estimated with this method differs significantly from the distribution expected for Gaussian noise. This is because a large number of high SNR signals are present in the data, which sometimes overlap for a given value of the time shift, generating partially coherent triggers with a larger-than-usual value of the detection statistic. This effectively spoils the sensitivity of the search by increasing the threshold on $p_\Lambda$ corresponding to a given FAR.
In order to correctly estimate the background distribution, we therefore use the \textit{null stream} instead: in a network of GW detectors with a closed configuration, such as the triangular configuration of ET, the sum of detector responses to any GW signal is zero, which provides a signal-free channel that can be used to estimate the properties of the noise \cite{NullStream}. When using this channel to estimate the background, we find a distribution compatible with Gaussian noise.

\subsection{Detection of compact binary coalescences}

\begin{figure}
\centerline{\includegraphics[width=0.7\linewidth]{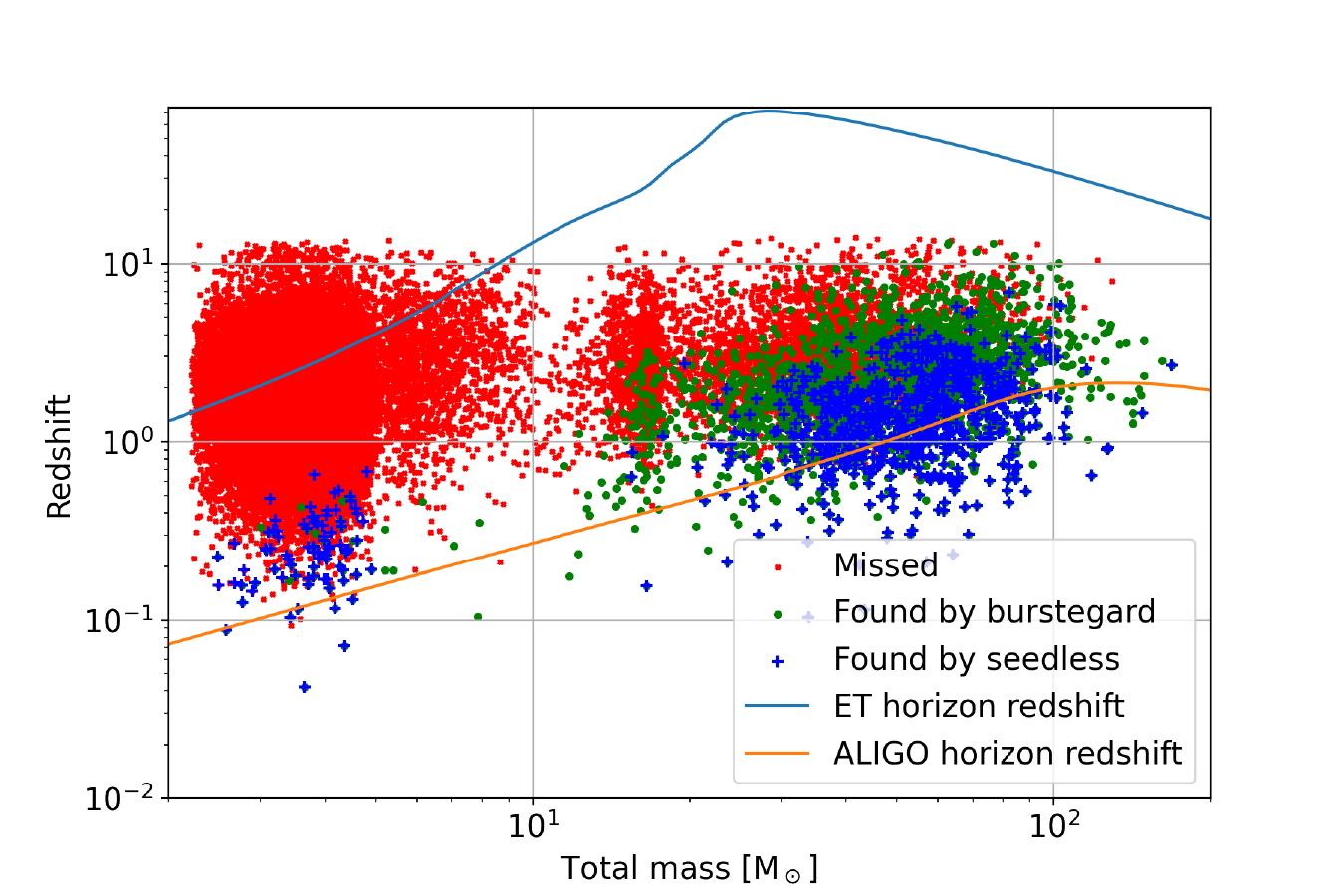}}
    \caption{Redshift of injected CBC signals as a function of their total mass in the source frame. Green dots represent signals recovered using the seed-based  \texttt{burstegard} clustering algorithm, while blue crosses are signals recovered by the seedless algorithm. Red crosses represent missed injections. The blue and orange curves representr the maximum redshift at which an equal-mass, non-spinning, quasicircular binary coalescence would be detectable with an optimal SNR $\geq 8$ in ET and Advanced LIGO at design sensitivity respectively.}
    \label{fig:M_vs_z}
\end{figure}

We compare the list of triggers extracted by the search to the list of signals injected in the data and find that $2656$ signals are recovered with a FAR lower than $1$ per year, including $2567$ BBH, $81$ BNS, and $8$ NSBH. In total, $38\%$ of the total number of BBH signals present in the data are recovered, up to a redshift $z\simeq 11$, and almost $90\%$ of the BBH with a total mass larger than $100$ M$_\odot$. 
The search is less sensitive to lower-mass signals, but the dedicated search algorithm used to target BNS is able detect a rate of $\sim 3$ BNS per day, and to provide an unbiased estimation of their chirp mass with a precision of $1\%$. The distribution of recovered and missed signal as a function of their total mass and redshift is shown in Fig. \ref{fig:M_vs_z}.

%\begin{figure}
%\centerline{\includegraphics[width=0.7\linewidth]{background}}
%\caption[]{Cumulative rate of noise triggers as a function of the detection statistic %$p_{\Lambda}$ for the two methods used to estimate the background.}
%\label{fig:bkg}
%\end{figure}

\section{Conclusion}

We show that \texttt{PySTAMPAS} is able to detect a significant number of CBC signals in ET for a very modest computational cost ($\sim 1$ day using off-the-shelf computing resources), and to accurately and rapidly estimate the chirp mass of the BNS signals detected, which could prove useful in the context of multi-messenger astronomy.
We also find that the large amount of high SNR signals in the data leads to overestimate the rate of noise triggers when estimated with the standard time-slides method, effectively spoiling the sensitivity of the search. That issue can be solved by using the null stream, which can be constructed by summing the response of the $3$ detectors in a triangular configuration, and provides a signal-free channel to estimate the background.

Although many CBC signals are recovered, the detection horizon of this search is still below what could be achieved with a matched-filter based search algorithm. The implementation of such search and comparison with what has been found here will be the focus of future work.

\section*{Acknowledgments}
Part of our simulations were performed on the Virtual Data cloud computing system at IJCLab.


\section*{References}

\begin{thebibliography}{99}

\bibitem{ET}M. Punturo {\it et al}, \Journal{Class. Quant. Gravity}{27}{194002}{2010}.
\bibitem{BB}A. Abac {\it et al}, \Journal{https://arxiv.org/abs/2503.12263}{}{}{2025}.
\bibitem{pystampas}A. Macquet {\it et al}, \Journal{\PRD}{104}{102005}{2021}.
\bibitem{paper}A. Macquet {\it et al}, \Journal{\PRD}{111}{022002}{2025}.

\bibitem{GWTC-3}R. Abbott {\it et al}, \Journal{Phys. Rev X}{13}{011048}{2023}.
\bibitem{NullStream}Y. G\"ursel {\it et al}, \Journal{\PRD}{40}{3884}{1989}.


\end{thebibliography}
\end{document}